\documentclass[10pt,aps,prr,final,twocolumn,amsmath,amssymb,superscriptaddress,longbibliography]{revtex4-2}
\usepackage[pdftex, colorlinks=true, linkcolor=blue, citecolor=blue, urlcolor=blue]{hyperref}
\usepackage{graphicx}
\usepackage{aurical}
\usepackage[T1]{fontenc}
\usepackage{units}
\usepackage[version=3]{mhchem}
\usepackage{tikz}
\usepackage{soul}
\graphicspath{{./figures/}}
\DeclareGraphicsExtensions{.png,.pdf}

\newcommand{\dif}{\mathrm{d}}

\newcommand{\una}{Universit\'{e} de Nouakchott, Facult\'{e} des Sciences et Techniques, D\'{e}partement de Physique,
	   Nouakchott, Mauritanie}

\newcommand{\br}{\mathbf{r}}

\newcommand{\VT}{V_{\rm T}}
\newcommand{\EF}{\varepsilon_{\rm F}}

\newcommand{\comment}[1]{}

\newcommand{\vt}{v_{\rm T}}
\newcommand{\rt}{\br_{\rm T}}
\newcommand{\Gr}{\mathcal G ^{(0)}(\br,\rt ,\varepsilon)}
\newcommand{\Grt}{\mathcal G^{(0)}(\rt,\rt ,\varepsilon)}
\newcommand{\wfa}{\psi _{l\varepsilon a}}
\newcommand{\wfaa}{\psi _{l\varepsilon a}^{(0)}}
\newcommand{\At}{\beta_{\varepsilon} (\rt)} 
\def\bepsilon{\bar{\varepsilon}}
\def\bbepsilon{\bar {\bar{\varepsilon}}}
\def\bk{\bar{k}}

\def\ba{\bar{a}}
\def\bba{\bar {\bar{a}}}
\def\bl{\bar{l}}
\def\bbl{\bar { \bar{l}}}
\def\tepsilon{\varepsilon^{(\mathrm{t})}}
\def\Me{M_{\rm e}}
\def\r]{\right]}
\def\l[{\left[}

\usepackage{amsmath}

\begin{document}
	\title{Scanning Gate Microscopy response for local tip potentials beyond perturbation theory}
	\author{Ousmane Ly}
	\email{ousmanebouneoumar@gmail.com}
	\affiliation{\una}
	
\begin{abstract}
	We propose an analytical formulation for the Scanning Gate Microscopy (SGM) response to local tips with arbitrary strength in two terminal nanostructures. 
	The real space resolved conductance is expressed in terms of the unperturbed quantities underlying the scattering problem. 
	Providing a non-dynamical approach for obtaining the SGM maps, the proposed expression enables for a significant reduction in the computational cost of SGM response calculations. This feature is particularly advantageous for deep learning-based approaches which have been recently proposed for  accessing local properties and disorder landscapes from conductance measurements. This opens up new possibilities for the SGM technique and holds exciting prospects for quantum transport.
	Further, the formula's versatility extends beyond this specific application, offering a straightforward and computationally efficient method for obtaining the SGM response in a more general context.
\end{abstract}

\maketitle
\section{Introduction}
Scanning gate microscopy is an experimental technique used to study space-resolved quantum transport features \cite{topinka2000, topinka2001, topinka2002}. In this technique, an atomic force microscopy (AFM) tip is capacitively coupled to a two-dimensional electron gas located at a certain distance from the scanned surface, and the conductance of the system is measured as the tip is moved throughout the structure. Conductance maps obtained through SGM have been found to display interesting local properties, such as branching flow of electrons out of quantum point contacts, wavefunction-related features in quantum rings and mesoscopic cavities \cite{sellier2011,  pala2009, cabosart2014}.

{It is important to note that the description of SGM might need to be 
	adapted for different setups, as the effect of the probing tip differs depending on the investigated systems. For instance in the case of confined geometries \cite{ Steinacher2016, Kozikov2016} or small sized quantum dots \cite{Michael2002, Pioda2004, Fallahi2005, Bleszynski2007, Schnez2011} the analysis of the impact of the tip's electrostatic potential turns out to be more subtle, as charging effects shall be taken into account. However, in the case of large systems these effects can be ignored and the analysis therefore becomes simpler. 
}

It was only more than a decade after the proposition of this experimental technique \cite{eriksson1996, topinka2000}, that an analytical formulation was proposed to describe the space resolved features underlying SGM. This was first worked out using a perturbation theory framework of the underlying scattering problem  and applied in the context of transport calculations in disorder-free quantum point contacts \cite{jalabert2010, gorini2013}. Further, the proposed perturbation theory has been utilized to establish a quantitative correspondence relation between SGM responses and partial local density of states (PLDOS) within the neighborhood of a quantum point contact in the presence of disorder \cite{Ly2017}. {This analytical derivation of the two lowest terms in tip strength of the SGM response has further triggered interesting efforts \cite{steinacher2017} towards the establishment of more sophisticated experimental setups bridging the commonly explored strongly invasive regime (non-perturbative) to the case where the tip strength is weakly invasive (perturbative). 

	While the perturbative treatment has been widely explored theoretically, 
	it remains of foremost importance to formulate a more general theoretical framework}
to describe the response to an arbitrary tip strength, at least in special cases like the one of a local probe that we treat in this work. Re-summing a perturbative series to infinite order is a notoriously difficult problem that can only be achieved in specific cases. Among them, the recently obtained zero SGM conductance correction for the peculiar zero transverse energy mode of a metallic armchair graphene nanoribbon under a long range tip potential \cite{Chen2023}.

Another motivation for developing the analytical approach of SGM beyond perturbation theory is to avoid the costly implementation of numerical schemes when a full transmission calculation should be performed for each tip position. Such a situation is encountered in the training of machine-learning based algorithms where a huge amount of SGM maps is required, considering different parameters such as the structure's size, the tip potential strength and temperature. This limitation has been recently pointed out in studies \cite{percebois2021, daCunha2022} that use machine learning techniques to extract key local transport features, namely PLDOS and disorder background in two-dimensional systems.

In the present work, we undertake the challenging task of deriving an analytical formula for computing SGM responses. The proposed formula assumes only a delta-like SGM tip with arbitrary strength and without restrictions on the considered geometries, which can be disordered and/or of any shape. The delta-likeness is achieved using short-range tip potentials whose real-space extent is below the Fermi wavelength. This is the regime where the SGM-PLDOS correspondence is indeed expected \cite{Ly2017}, although under very moderate strengths of the probing tip. Interestingly, the full SGM response can be systematically deduced from the perturbation theory results. We find that the SGM conductance of the structure is simply related to the system's scattering matrix, its unperturbed scattering wave-functions and the real-space diagonal elements of the retarded Green's function. 

\section{Analytical Formulation of SGM in the Perturbative Regime}
To derive the exact formula for the  SGM response, we find it pedagogical to first recall the lowest order terms of the perturbation theory for the conductance corrections due to a local potential, as described in Refs. \cite{jalabert2010, gorini2013}. Further, we demonstrate that the entire SGM response series can be deduced straightforwardly from these perturbative results through a mere renormalization of the tip potential matrix elements.

To this end, we consider an arbitrary quantum scatterer, located at position $x=0$ and attached to two semi-infinite leads. The asymptotic form of the unperturbed scattering wave-functions (in the absence of the tip potential) can be written in terms of the elements of the scattering matrix and the wave-functions ($\varphi_{1\varepsilon a}^{(\pm)}$) of the quasi one-dimensional free electron leads:

\begin{align}\label{allscats}
\psi_{1\varepsilon a}^{(0)}(\br) 
&= 
\left\lbrace 
\begin{array}{ll}
\varphi_{1\varepsilon a}^{(-)}(\br) + \sum_{b=1}^{N} r_{ba} \, \varphi_{1\varepsilon a}^{(+)}(\br),
& x <0 \\
\sum_{b=1}^{N} t_{ba} \, \varphi_{2\varepsilon b}^{(+)}(\br), & x >0  
\end{array} 
\right.  \nonumber \\
\psi_{2\varepsilon a}^{(0)}(\br) 
&= 
\left\lbrace 
\begin{array}{ll}
\sum_{b=1}^{N} t'_{ba} \, \varphi_{1\varepsilon a}^{(+)}(\br),
& x <0 \\
\varphi_{2\varepsilon a}^{(-)}(\br) +\sum_{b=1}^{N} r'_{ba} \, \varphi_{2\varepsilon b}^{(+)}(\br), & x >0  
\end{array} 
\right. \nonumber \\ 
& &
\end{align}

Here, the $\pm$ signs on the leads' wavefunctions denote, respectively, outgoing and impinging energy modes,  and $l=1,2$ stands for the lead number. The matrices $r$, $r'$, $t$, and $t'$ are elements of the scattering matrix $S$ defined as:

\begin{equation} \label{S}
S = \left( \begin{array}{cc}
r & t' \\
t & r'
\end{array} \right).
\end{equation} 

Next, we consider a potential $V_T(\br)$ that perturbs the initial system described above. The underlying scattering wave-functions in Eq. \eqref{allscats} are therefore modified.

Up to linear order in the tip potential, the correction to the scattering wave-function in the presence of the tip is obtained using the  Lippmann Schwinger expansion \eqref{eq:Born_approx_n}.

Assuming a delta tip potential $V_{\rm T}(\br)=v_{\rm T}\delta(\br - \rt)$,  the first order correction to the scattering wave function reads
\begin{equation}\label{eq:psi-first}
\psi_{l,\varepsilon,a}(\br)=\psi_{l,\varepsilon,a}^{(0)}(\br)+v_{\rm T} \mathcal{G}^{(0)}(\br,{\rt},\varepsilon)  
 \psi_{l,\varepsilon,a}^{(0)}({\rt}),
\end{equation}
with $\mathcal{G}^{(0)}$ being the unperturbed retarded Green function. 
The two lowest order corrections to the conductance are obtained as (See Appendix. \ref{ap:perturbationsgm})

\begin{equation}
\label{eq:g11-main}
g^{(1)} = 4 \pi \mathrm{Im}\left\lbrace \mathrm{Tr}\l[ 
t^{\dagger}t\ \mathcal{V}^{11}
+t^{\dagger}r'\ \mathcal{V}^{21}\r]
\right\rbrace ,
\end{equation}
and
\begin{equation}
\label{eq:g2beta-main}
{g^{(2)} = 4\pi^2  \mathrm{Tr}\left\lbrace 
	{\rm {Re}}[\mathcal{V}^{11}(t^\dagger t \mathcal{V}^{11}+2t^\dagger r' \mathcal{V}^{21}+r'^\dagger r'\mathcal{V}^{22})]
	\right\rbrace .}
\end{equation}

It can be observed that the first term in the rectangular bracket of Eq. \eqref{eq:g11-main} is real. Therefore, the conductance correction can be further simplified, and only the second term would survive after taking the imaginary part. In the actual context, we consider the most general expression as the formula will be applied to complex matrix elements as we will see in the following. 

\section{Higher Order Terms of the Conductance Expansion}
In order to calculate the higher order terms of the scattering wave-function, we start again from the Lippmann-Schwinger expansion applied to a delta tip potential. 
The resulting corrected scattering wave-function reads

\begin{equation}\label{eq:lip-delta}
\wfa (\br)=\wfa ^{(0)}(\br)+\vt \Gr \wfa (\rt).
\end{equation}

To obtain the above scattering wave-function only in terms of the unperturbed quantities, we first evaluate $\wfa (\br)$ at the tip position $\rt$ 

\begin{equation}\label{lipmannshort}
\wfa (\rt) =\At\wfa ^{(0)}(\rt),
\end{equation}
where $\At$ is a tip position dependent complex function defined as

\begin{equation}\label{beta}
\At =1/(1-\vt\Grt).
\end{equation}

By plugging \eqref{lipmannshort} into \eqref{eq:lip-delta}, we find that the scattering state at an arbitrary position $\br$ is simply given by

\begin{equation}\label{psi}
\wfa (\br)=\wfaa (\br) +\vt\At \Gr \wfaa (\rt).
\end{equation}

This expresses the scattering wave-function in terms of the unperturbed quantities. That is in the absence of the perturbing probe. 
It can be noticed that Eq. \eqref{psi} is equivalent to Eq. \eqref{eq:psi-first}, up to the prefactor $\At$ in the second term on the right-hand side of the latter. This is the key aspect that allows for a simple deduction of the full conductance correction corresponding to \eqref{psi} from the previous perturbation theory results. The generalized conductance corrections can be straightforwardly obtained from \eqref{eq:g11-main} and \eqref{eq:g2beta-main}. The prefactor $\At$ would simply alter the matrix elements $\mathcal{V}^{ll'}$, leading to a complex factor $\At$ at the level of the first-order like contribution, and $|\At|^2$ at the level of the second order-like correction.
Therefore, we are left with the following expression

\begin{widetext}
\begin{equation}\label{fullg}
{g = 4\pi {\rm{Im}} \{\At {\rm {Tr}}[t^\dagger t \mathcal{V}^{11}+t^\dagger r' \mathcal{V}^{21}]\} 
	+4\pi^2 |\At|^2 { \mathrm{Tr}\left\lbrace 
		{\rm {Re}}[\mathcal{V}^{11}(t^\dagger t \mathcal{V}^{11}+2t^\dagger r' \mathcal{V}^{21}+r'^\dagger r'\mathcal{V}^{22})]
		\right\rbrace}.}
\end{equation}
\end{widetext}
Equation \eqref{fullg} is the central result of this work. It gives the analytical SGM response in the presence of a delta-like probe with arbitrary tip strength. The formula does not assume any particular shape of the scattering region, which may also include disorder. Although the formula is expressed in terms of the two lowest orders of the perturbation theory, the resulting conductance is neither linear nor quadratic in $v_{\rm T}$. In fact, the complex coefficient $\At$ contains higher-order terms, as it is expressed as a power series of $v_{\rm T}\mathcal{G}^{(0)}$. 

\section{Numerical implementation}
In order to validate the proposed full SGM response Eq. \eqref{fullg}, we performed numerical simulations on a disordered ring geometry defined on a tight binding network with lattice parameter $a$. The ring had an inner radius of $25a$ and an outer radius of $50a$. The underlying momentum dependent Hamiltonian is given by 
\begin{equation}\label{eq:ham}
H = \frac{1}{2m^*}(k_x^2+k_y^2) + V_d(\br) + V_{conf}(\br),
\end{equation}
where $k_x$ and $k_y$ are the momenta in the two dimensions of space. The terms $V_d$ and  $ V_{conf}$ stand respectively for disorder and confining potentials. To compute the conductance of the system the ring is attached to two free electron leads.
The effective mass was taken as $m^*=0.04 m$, where $m$ is the bare electron mass as in Ref. \cite{pala2009}. 
The Hamiltonian Eq. \eqref{eq:ham}, is discretized on the tight-binding lattice. Subsequently, the scattering problem is solved using the quantum transport package Kwant \cite{kwant}.

\begin{figure*}[htbp]
	\centering
	\includegraphics[width=.8\textwidth]{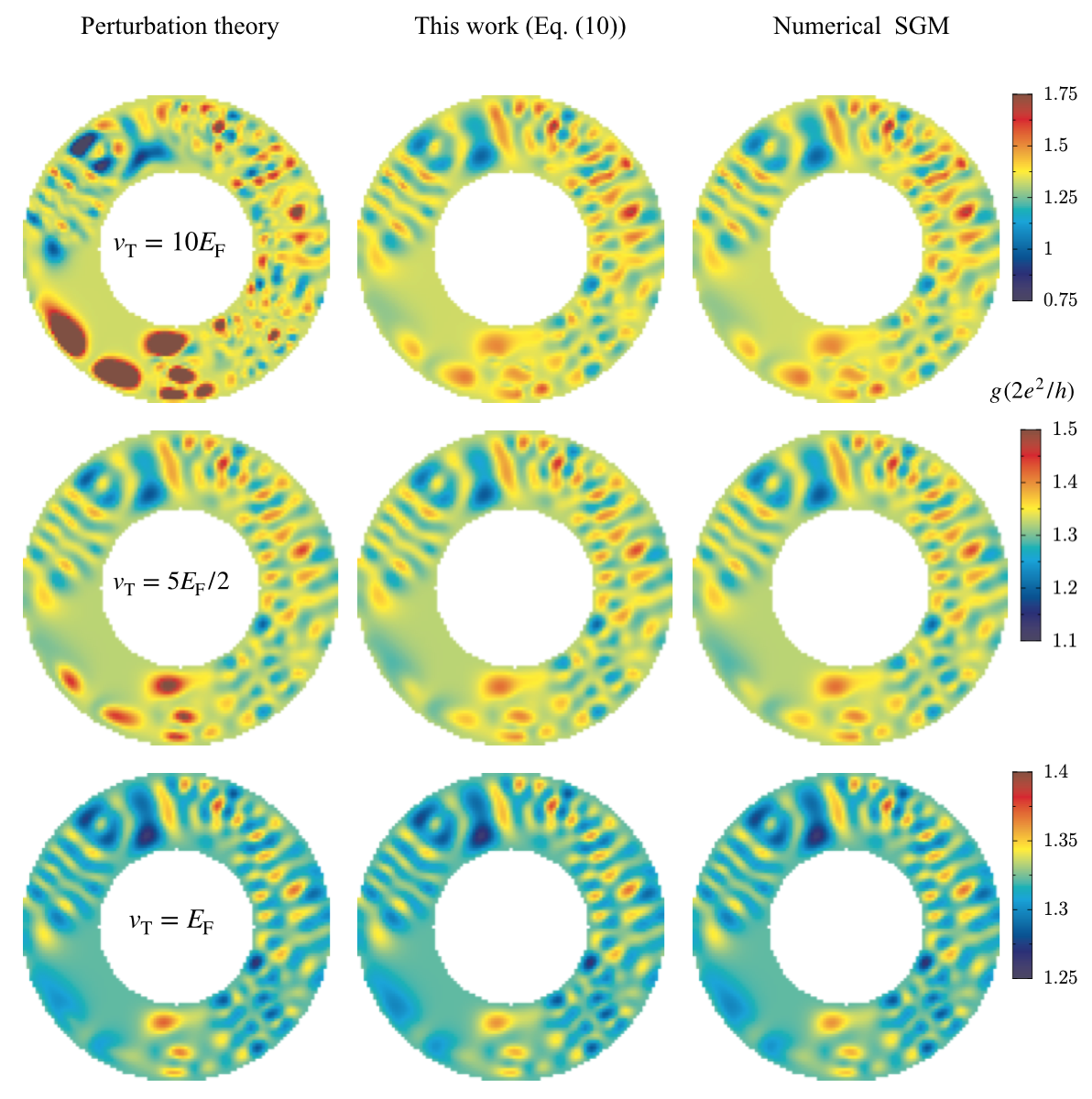}
	\caption{SGM response at different tip potentials is shown. The analytical formula of this work (middle panels) is compared to the exact numerical simulations (right panels). We also display the perturbation theory \cite{jalabert2010} result in the left column. One can observe the deviation of the perturbative results from the exact numerical calculations as $v_t$ is increased. While the analytical formula (Eq. \eqref{fullg}) reproduces nicely  the exactly numerically calculated SGM responses.}
	\label{fig:maps}
\end{figure*}

In Fig. \ref{fig:maps}, we plotted the real space resolved conductance at different tip potentials. The exact numerical simulations are displayed in the right column. In the central panels, the results corresponding to the implementation of the analytical formula (Eq. \eqref{fullg}) are shown. In the left column, the lowest order conductance correction of the perturbation theory \cite{jalabert2010} is computed. As expected, the perturbative results remain a good approximation of the response at low $v_t$. However, the analytical formula reproduces nicely the full numerical calculations at all considered tip strengths.

\begin{figure}
	\includegraphics[width=\linewidth]{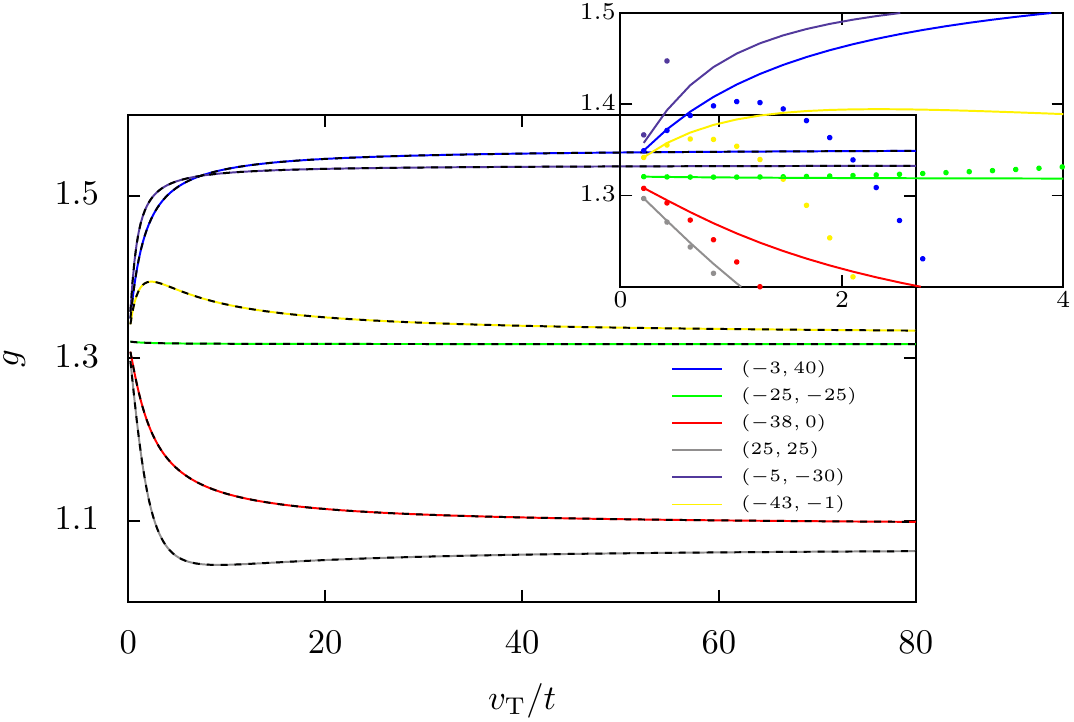}
	\caption{
		The conductance of the quantum ring is plotted as a function of the strength of a local tip.  Each color line corresponds to an arbitrarily chosen tip position in the ring geometry. The dashed lines correspond to the analytical prediction (Eq. \eqref{fullg}).  In the inset  the sum of the two lowest order  corrections  to $g$  (dots) is compared to the full SGM conductance (solid lines) given by Eq. \eqref{fullg}.}	
	\label{saturation}
\end{figure}

Furthermore, we focus on the high tip voltage limit, as this is the regime where the perturbation theory breaks down.
In Fig. \ref{saturation}, we plotted the conductance of the ring versus the strength $\vt$ of the local tip. Each color line corresponds to an arbitrarily chosen tip position in the disordered ring. The dashed lines correspond to the analytical prediction (Eq. \eqref{fullg}). The full SGM response obtained by means of the exact numerical computation is represented by the solid lines. 
It can be noticed that the SGM response at the considered positions behaves differently. 
Semi-classically speaking, some tip locations might favor the reflection of impinging trajectories, while others would only slightly deviate them. The former case would therefore favor a reduction of the transmission with respect to its unperturbed value, leading to a decrease in the computed conductance. In the latter scenario, the presence of the tip might trigger additional trajectories that are unexpected in its absence and therefore lead to the enhancement of the conductance with respect to its value in the unperturbed structure. One observes that the analytical predictions coincide perfectly with the fully numerically calculated SGM response. This provides a validation of the proposed formula. Yet, we shall emphasize that the perturbation theory fails drastically at reproducing the exact SGM conductance, though a delta tip was assumed.

Although experimentally extended tips are often used in SGM, the delta-tip conductance formula effectively captures all the relevant features. It is important to note that in the presence of an extended tip, the electrostatic profile would blur the corresponding local responses and trigger a stronger response, which scales with the diameter of the probe. 

\begin{figure*}
	\includegraphics[width=\linewidth]{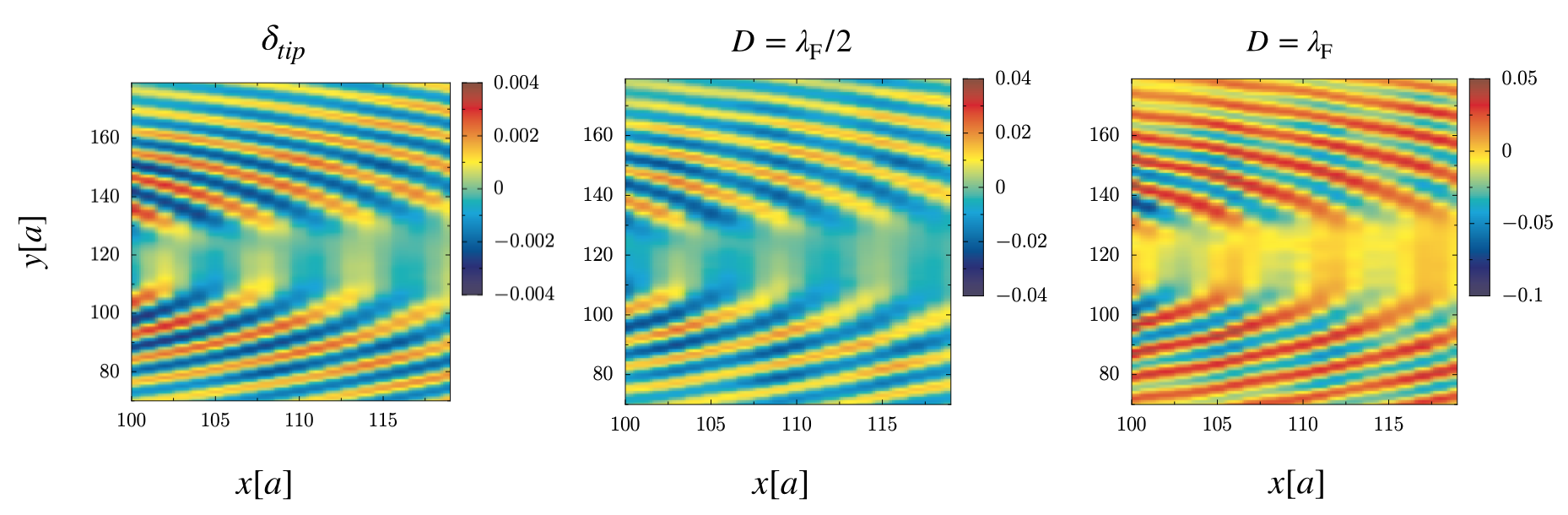}
	\caption{The SGM response of a quantum point contact located at the center-left of each map. Different diameters of the hard disc potential are considered. The tip strength is set at $v_{\rm T}=E_{\rm F}$ and the unperturbed transmission is tuned at $g=1.82$. The color axes show the change of conductance induced by the tip perturbation.}
	\label{fig:diameter}
\end{figure*}

{In order to investigate the impact of increasing the tip size on SGM, we consider, without loss of generality, the case of an abrupt quantum point contact (QPC). 
	
	Certainly, the nature of the studied geometry plays a role. In fact, in the case of a quantum ring such as the one considered above, placing a large scanning disc would lead to a strong reduction of conductance while a $\delta-$tip potential allows for larger transmission and only reflects a part of the impinging flux passing by the position where the local tip is placed. This leads obviousely to an accentiated difference between extended and local tip responses due to the particular geometry. However, for a large enough quantum ring, whose all relevent dimentions are larger than the tip size, both  local and extended tip responses are expected to be well correlated. 
	
	In Fig. \ref{fig:diameter}, the SGM response of a delta-tip (left panel) is compared to the response of an extended tip with diameter $\lambda_{\rm F}/2$ ($\lambda_{\rm F}$) in the central (right) panels.}
While the main features of the SGM response are clearly captured by the delta-tip potential, the extended tip responses appear to be stronger in amplitude. {Note that the tip strength is kept constant whereas the disc diameter is varied. 
} 
In these simulations a weak Anderson like disorder was considered leading to the asymmetry of the obtained maps.

\section{Discussion }
The proposed analytical formula for the real space resolved conductance in the presence of a local tip  is expressed in terms of the elements of the scattering matrix, the scattering states, and the unperturbed Green's function. An important simplification arises from the fact that these quantities only need to be computed once, even if the tip strength is modified. However, calculating the real-space Green's function throughout the studied geometry can be memory consuming for very large systems, if the brute force diagonalization method is employed, and caution should be taken in this case. In our simulations, we used the Kwant software \cite{kwant} to obtain the real-space retarded Green's function, but further theoretical efforts have been devoted to make the computation of $\mathcal{G}_0$ more efficient for large systems \cite{istasthesis}. Moreover, evaluating the Green's function by directly computing the underlying energy integrals is also feasible using the non-equilibrium transport package Tkwant \cite{tkwant}, which also incorporates the Green's function formalism.

The actual analytical formulation can be particularly useful in machine learning-based approaches designed to obtain local properties within an inverse problem framework. These approaches require a large number of SGM maps to train the underlying neural networks. A task that would be impossible to achieve or very demanding if the standard numerical procedure of computing SGM is followed. The formula we propose enables for the optimization of these calculations and further facilitates the incorporation of pertinent parameters of the problem, such as the system's size and the tip strength in addition to temperature, with a minimal computational expense. 
We believe that our proposal will enable new possibilities in harnessing SGM data to provide unprecedented access to quantum transport properties in mesoscopic systems. Furthermore, our analytical formulation can be applied in the study of nonlinear quantum transport \cite{gorini2014} and SGM induced thermoelectric effect \cite{abboutthesis,brun2019,fleury2021}. 

{
In the extended tip simulations an effective hard disc has been used to model the SGM response. Although the tip profile is often modeled using a Lorantzian or a Gaussian, it is admitted \cite{Ly2017} that the hard disc model remains a good approximation of these tip profiles. And leads to similar features regardless of the way the tip profile decays exactly.
	Further, it is worth emphasizing that our proposal might not be adapted to fully describe the response to an extended electrostatic potential with arbitrary strength. However, it gives an interesting perspective for addressing SGM experiments in the strongly invasive regime, where the tip-induced potential is much larger than the Fermi energy of the underlying system.
	So far, the proposed approach remains of great advantage even for extended tip potentials, in the situation where transport features are extracted through machine learning techniques. Indeed, one can assume that any given SGM image would be mapped to some auxiliary $\delta-$tip response. Accordingly, it would still be a good idea to train the corresponding neural networks using the conductance expression given by Eq. \eqref{fullg}. Nonetheless, this raises a question of the uniqueness of the SGM-PLDOS and SGM-disorder correspondence relations. 
	This is on one hand. On the other hand,
the $\delta-$tip SGM - disorder pairs computed through our formalism can be utilized in the context of the transfered learning approach recently employed in ref. \cite{Percebois2023} to train a neural network with the easy to calculate PLDOS-disorder pairs, with the aim to extract disorder configurations from experimental SGM responses in a high-mobility semiconductor heterostructure hosting a quantum point contact.
In this interesting approach the learning mechanism has led to more than $80$\% accuracy in the extraction of the real space potential configuration. Indeed, we would expect the precision of the extraction method to be rather enhanced if the $\delta-$tip SGM expression is implemented instead of PLDOS. Although the latter is often found to look very similar to the SGM response, especially in particular systems such as quantum point contacts, the former appears to be much more correlated with the finite tip size conductance calculations in various geometries. For instance, the real space coherent fringes characteristic of SGM are not always present in the partial local density of states. In contrast, these real space resolved features are common to both $\delta-$like and extended tip profile responses (see Fig. \ref{fig:diameter}, e.g.), in both clean and disordered landscapes.}

Finally, we highlight that experimental efforts have been devoted to fabricate shielded co-axial tips \cite{Harjee2010, Harjeethesis}. However, the derivation of a more general formula remains of relevant interest to more quantitatively describe experiments in the extended tip regime.  
In addition to the above mentioned applications in machine learning based approaches, the actual formalism will be certainly insightful for approaching this interesting task, that can be addressed at least in specific albeit relevant circumstances. 

\section{Conclusion}
In summary, 
we have demonstrated 
that the full SGM response series can be obtained in an exact fashion, provided that a very narrow scanning tip is assumed. 
By numerically implementing the formula, one can obtain SGM maps at a minimal computational cost. 
The analytical formula is demonstrated to be in perfect agreement with exact numerical evaluation of the SGM response. The possibility of applying the obtained SGM expression in the context of machine learning based approaches has been discussed.

\acknowledgments
We thank D. Weinmann and R. Jalabert for their careful reading of the manuscript, and for helpful discussions and suggestions. We also thank T. Ihn, X. Waintal and A. Abbout for useful discussions. 
We are grateful to the hospitality of IPCMS and University of Strasbourg where this work was initiated.

\newpage
\appendix

\section{Evaluation of the energy integrals involved in the corrections to the current density}
\label{energyintegrals}


When calculating the conductance corrections, we encounter the following energy integrals:


\begin{equation}\label{i1sappa}
	I_1^s(\varepsilon)= \int_{\varepsilon^{(t)}_1}^{\infty}\frac{\dif \bar{\varepsilon}}{\varepsilon^+-{\bbepsilon}}F({{\bbepsilon}})e^{si\bar{\bar{k}}x},
\end{equation}
and
\begin{equation}\label{i2sappa}
	I_2^{(s,\sigma)}(\varepsilon)= \int_{\varepsilon^{(t)}_1}^{\infty}\frac{\dif \bar{\varepsilon}}{\varepsilon^--\bar{\varepsilon}} 
	\frac{\dif \bbepsilon}{\varepsilon^+-\bar{\varepsilon}}
	H({\bar{\varepsilon}},\bbepsilon)e^{i(\sigma\bar{k}+s\bar{\bar{k}})x},
\end{equation}
where a smooth dependence of $F$ and $H$ on the energy variables is assumed.

Changing the energy variables to momenta, the integrals become
\begin{equation}
	I_1^s(\varepsilon)= \int_{0}^{\infty}\frac{-2\bar{\bar{k}}\dif \bar{\bar{k}}}{[\bar{\bar{k}}+k+i\bar{\eta}][\bar{\bar{k}}-(k+i\bar{\eta})]}F({\bar{\bar{\varepsilon}}})e^{si\bar{\bar{k}}x},
\end{equation}
and 
\begin{widetext}
\begin{equation}
	I_2^{(s,\sigma)}(\varepsilon)= \int_{0}^{\infty}\frac{-2\bar{k}\dif \bar{k}}{[\bar{k}+k+i\bar{\eta}][\bar{k}-(k-i\bar{\eta})]} 
	\frac{-2\bar{\bar{k}}\dif \bar{\bar{k}}}{[\bar{\bar{k}}+k+i\bar{\eta}][\bar{\bar{k}}-(k+i\bar{\eta})]}
	H({\bar{\varepsilon}},\bbepsilon)e^{i(\sigma\bar{k}+s \bar{\bar{k}})x}.
\end{equation}
\end{widetext}

To evaluate the energy integral  $I_1^s$ we use one of the two quadrants of the complex plane depicted in Fig. \ref{fig:c1}.  If $s=-1$, it is appropriate to chose $Q_u$, in order to ensure that (i) the integration along the corresponding quarter vanishes as $|\bar{k}|$ tends to $\infty$; and (ii) the contribution along the imaginary axis vanishes in the limit of  $x$ tending to infinity. 
In contrast if $s=+$, $Q_d$ should be considered.

Since there is no pole in $Q_d$, we have $I_1^-=0$, according to the residue theorem.
On the other hand the pole $k+i\bar{\eta}$ lies in $Q_u$, therefore the application of the residue theorem gives 
\begin{equation}\label{i1pappa}
	I_1^+(\varepsilon)=-2\pi iF(\varepsilon)e^{ikx}.
\end{equation}
Similarly, the $\sigma$-integral of $I_2$ (the $\bar{k}$ integral) vanishes for $\sigma=+$, since there is no pole in $Q_u$ (see Fig. \ref{fig:c2}). Therefore, $I_2$ vanishes if $\sigma=+$ and/or $s=-$.
However, if $\sigma=-1$ and $s=1$ the residue theorem gives
\begin{equation}\label{i2pmappa}
	I_2^{(+,-)}(\varepsilon)=4\pi ^2 H(\varepsilon,\varepsilon).
\end{equation}

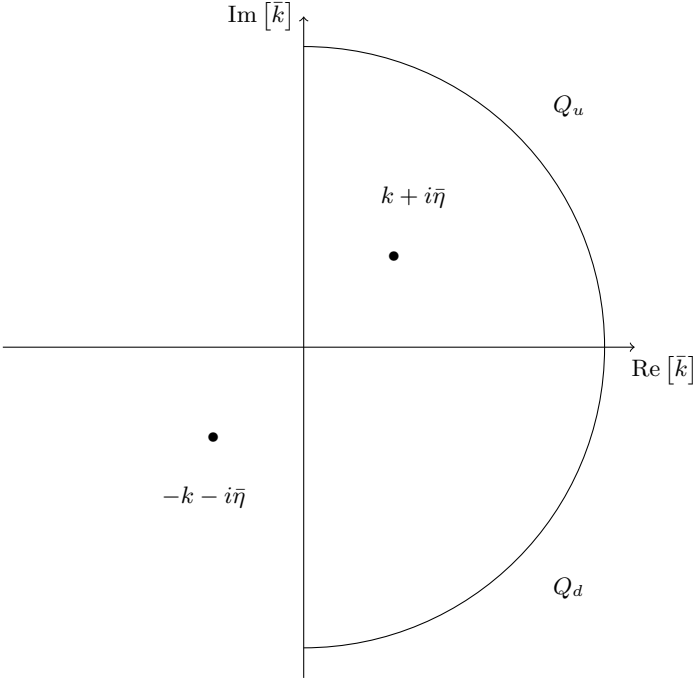
\begin{figure}
	\centering
	\begin{tikzpicture}
	\draw  [scale=4][->] (-1,0) -- (1.1,0); 
	\draw [scale=4] (1.2,0) node[below] {$\mathrm{Re}\l[\bar{k}\r]$}; 
	\draw [scale=4] [->] (0,-1.1) -- (0,1.1); 
	\draw [scale=4](0,1.1) node[left] {$\mathrm{Im}\l[\bar{k}\r]$}; 
	\draw [scale=4](1,0) arc (0:90:1) ;
	\draw [scale=4](1,0) arc (0:-90:1) ;
	\draw [scale=4](0.5,0.5) node[left] {$k+i\bar{\eta}$};
	\draw [scale=4](0.3,0.3) node {$\bullet$} ;
	\draw [scale=4](-0.5,-0.5) node[right] {$-k-i\bar{\eta}$};
	\draw [scale=4](-0.3,-0.3) node {$\bullet$} ;
	
	\draw [scale=4](0.8,0.8) node[right] {$Q_u$};
	\draw [scale=4](0.8,-0.8) node[right] {$Q_d$};
	\end{tikzpicture} 
	\caption{The complex plane of $\bar{\bar{k}}$ used to calculate $I_1$, and the $\bar{\bar{k}}$ integral of $I_2$ is shown. The orientation of the contours is dictated by the real axis orientation. The poles of the integral are represented by the dots.}
	\label{fig:c1}
\end{figure}
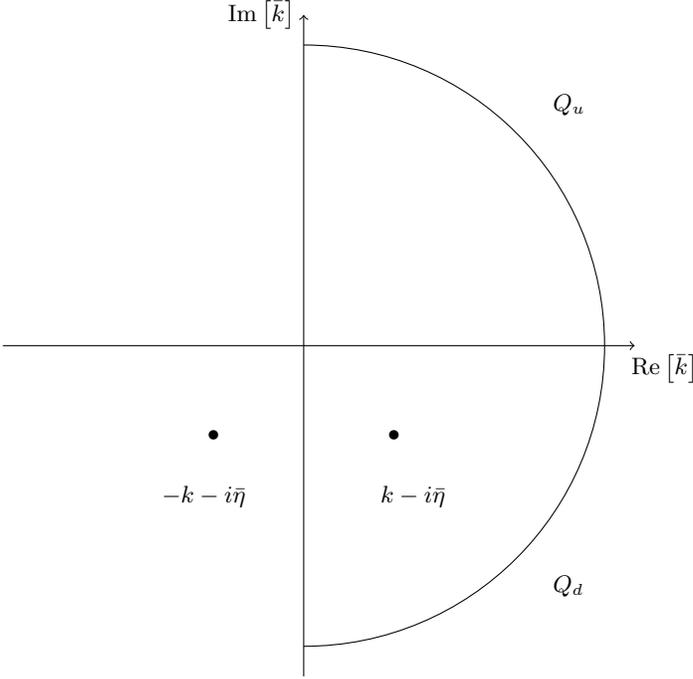
\begin{figure}
	\centering
	\begin{tikzpicture}
	\draw  [scale=4][->] (-1,0) -- (1.1,0); 
	\draw [scale=4] (1.2,0) node[below] {$\mathrm{Re}\l[\bar{k}\r]$}; 
	\draw [scale=4] [->] (0,-1.1) -- (0,1.1); 
	\draw [scale=4](0,1.1) node[left] {$\mathrm{Im}\l[\bar{k}\r]$}; 
	\draw [scale=4](1,0) arc (0:90:1) ;
	\draw [scale=4](1,0) arc (0:-90:1) ;
	\draw [scale=4](0.5,-0.5) node[left] {$k-i\bar{\eta}$};
	\draw [scale=4](0.3,-0.3) node {$\bullet$} ;
	\draw [scale=4](-0.5,-0.5) node[right] {$-k-i\bar{\eta}$};
	\draw [scale=4](-0.3,-0.3) node {$\bullet$} ;
	\draw [scale=4](0.8,0.8) node[right] {$Q_u$};
	\draw [scale=4](0.8,-0.8) node[right] {$Q_d$};
	\end{tikzpicture} 
	\caption{The complex plane of $\bar{k}$ used to calculate $I_2$ is shown. The orientation of the contours is dictated by the real axis orientation.}
	\label{fig:c2}
\end{figure}

\section{SGM response in the perturbative regime}
\label{ap:perturbationsgm}
To provide the elementary background for the derivation of the exact SGM response formula, we first derive the lowest order terms of the perturbation theory for the conductance corrections due to a local potential, following Ref.\cite{jalabert2010}. 

Let us consider a quantum scatterer attached to two semi infinite leads. The corresponding asymptotic form of  the unperturbed scattering wave-functions are given by Eq. ({\color{blue} 1}). 

In the presence of the perturbing tip, the $n^{th}$ order correction to the scattering wave-function is obtained from the Lippmann-Schwinger equation as

\begin{equation}\label{eq:Born_approx_n}
	\Psi_{l,\varepsilon,a}^{(n)}(\br)=
	\int \dif \bar{\br} \ \mathcal{G}^{(0)}(\br,\bar{\br},\varepsilon)  
	\VT(\bar{\br})  \Psi_{l,\varepsilon,a}^{(n-1)}(\br') \ ,
\end{equation}
where $\VT(\bar{\br})$ is the potential induced by the tip. And $\mathcal{G}^{(0)}$ is the unperturbed Green's function defined as  
\begin{equation}
	\label{eq:G0}
	{
		\mathcal{G}^{(0)}(\br,\bar{\br},\varepsilon) = \sum_{\bar{l}=1}^{2} 
		\int_{\varepsilon^{(\mathrm{t})}_1}^{\infty} 
		\frac{\dif\bar{\varepsilon}}{\varepsilon^{+}-\bar{\varepsilon}}
		\sum_{\bar{a}=1}^{\bar{N}} \ \psi_{\bar{l},\bar{\varepsilon},\bar{a}}^{(0)*}(\bar{\br}) 
		\psi_{\bar{l},\bar{\varepsilon},\bar{a}}^{(0)}(\br) \ }.
\end{equation} 
Up to linear order in the tip potential the  correction to the scattering wave-function in the presence of the tip is obtained by taking $n=1$ in (\ref{eq:Born_approx_n}), which gives
\begin{equation}\label{eq:Born_approx}
	\Psi_{l,\varepsilon,a}^{(1)}(\br)=
	\int \dif \bar{\br} \ \mathcal{G}^{(0)}(\br,\bar{\br},\varepsilon)  
	\VT(\bar{\br})  \Psi_{l,\varepsilon,a}^{(0)}(\bar{\br}) \ .
\end{equation}
Hence, it becomes straightforward to write the full scattering wave-function up to first order in the unperturbed wave-functions as in Eq. ({\color{blue} 3}). To shorten the notation we denote by  ${\Psi}_{l\varepsilon a} ^{(1)}(\br)$, the last term of the right hand side of Eq. ({\color{blue} 3}). Two corrections to the current density can be identified as first $\delta ^{(1)}J_{l\varepsilon a}$ and second $\delta ^{(2)}J_{l\varepsilon a}$ order contributions. They read

\begin{equation}\label{J1_0}
	\delta ^{(1)}J_{l\varepsilon a} = \frac{e\hbar}{M_e}{\rm{\rm{Im}}}({\Psi}_{l\varepsilon a} ^{(0)*}(\br)\partial_x {\Psi}_{l\varepsilon a} ^{(1)}(\br)-\Psi_{l\varepsilon a}^{(1)} (\br)  \partial_x  {\Psi}_{l\varepsilon a} ^{(0)*}(\br)),
\end{equation}
and 
\begin{equation}\label{J2beta1}
	\delta ^{(2)}J_{l\varepsilon a} = \frac{e\hbar}{M_e}{\rm{\rm{Im}}}({\Psi}_{l\varepsilon a} ^{(1)*}(\br)\partial_x {\Psi}_{l\varepsilon a} ^{(1)}(\br)).
\end{equation}

Note that the considered second order contribution is simply one term of the two contributions calculated in Refs. \cite{gorini2013, jalabert2010}. Indeed, the extra second order term does not enter into play in the calculation of the non-perturbative SGM response. The missing second order contribution would only appear if the Lippmann Schwinger expansion is taken to second order in $\VT$. That is, if the term $n=2$ of \eqref{eq:Born_approx_n} is included. Therefore, a contribution involving the cross product of $\psi^{(0)}$ and $\psi^{(2)}$ would appear and this corresponds to the omitted second order term. Nonetheless, it is clear that such a term would not appear in the current densities resulting from Eq. ({\color{blue} 3})  or Eq. ({\color{blue} 9}). In this sense the second order correction considered in this work should be approached with care. Such contribution (the omitted one) was dubbed $\alpha-$like in \cite{gorini2013,lythesis,szewcthesis} (see Eq. ({\color{blue}35a}) in \cite{gorini2013}).

Let us first deal with the first order contribution.
Substituting  ${\Psi}_{l\varepsilon a} ^{(1)}$  by its expression (\ref{eq:Born_approx}) in Eq. (\ref{J1_0}), the current correction up to first order in $\VT$ reads
\begin{widetext}
	\begin{equation}
		\label{eq:j1}
		\delta ^{(1)}J_{l\varepsilon a}(\br)
		= 2\sum_{\bar{l}=1}^{2} \mathrm{Re}\left\lbrace 
		\int_{\varepsilon^{(\mathrm{t})}_1}^{\infty} 
		\frac{\dif\bar{\varepsilon}}{\varepsilon^{+}-\bar{\varepsilon}} \
		\sum_{\bar{a}=1}^{\bar{N}} 
		\left[j(\br)\right]_{a \ba}^{l, \bl}(\varepsilon,\bar{\varepsilon}) \
		\left[\VT\right]^{\bl ,l}_{\ba a}(\bar{\varepsilon},\varepsilon) 
		\right\rbrace \, ,
	\end{equation}
\end{widetext}
where 
\begin{widetext}
	\begin{equation}\label{j0matrix}
		\l[j(\br)\r]_{ a\ba}^{l, \bl}(\varepsilon,\bar{\varepsilon})=
		\frac{e\hbar}{2i\Me}
		\l[
		\Psi_{l,\varepsilon,{a}}^{(0)*}(\br) 
		\ {\partial _x}\Psi_{\bl,\bepsilon,\bar{a}}^{(0)}(\br) -
		\Psi_{\bl,\bepsilon,\bar{a}}^{(0)}(\br) \ {\partial _x}\Psi_{l,
			{\varepsilon},{a}}^{(0)*}(\br) 
		\r],
	\end{equation}
\end{widetext}
is the longitudinal current density matrix element after integration over the transverse direction, 
and 
\begin{equation}\label{vmatrix}
	\l[\VT\r]_{\ba a}^{\bl, l}(\bar{\varepsilon},\varepsilon)=\int \Psi_{\bar{l},\bar{\varepsilon},\bar{a}}^{(0)*}(\bar{\br}) 
	\VT(\bar{\br})  \Psi_{l,\varepsilon,a}^{(0)}(\bar{\br}) \dif \bar{\br}
	,
\end{equation}
the matrix element of the tip potential between two scattering wave-functions.

If we are interested in the current coming from the left to the right we shall take $l=1$. 
\begin{widetext}
	\begin{eqnarray}\label{detailed1}
		\l[j(\br)\r]_{\ba a}^{11}(\varepsilon,\bar{\varepsilon}) &=& \frac{e}{2h} \sum_{b=1}^{\hat{N}}
		\left(\sqrt{\frac{\bar{k}_b}{k_b}}+
		\sqrt{\frac{k_b}{\bar{k}_b}}\right) 
		t^{*}_{ba}{\bar t}^{\phantom{*}}_{b\bar{a}}
		\exp{\left[i(\bar{k}_b-k_b)x\right] }
	\end{eqnarray}
	
	\begin{eqnarray}\label{detailed2}
		\l[j(\br)\r]_{\ba a}^{12}(\varepsilon,\bar{\varepsilon}) &=& 
		\frac{e}{2h} 
		\left(\sqrt{\frac{k_{\bar{a}}}{\bar{k}_{\bar{a}}}}-
		\sqrt{\frac{\bar{k}_{\bar{a}}}{k_{\bar{a}}}}\right)
		t^{*}_{\bar{a}a}\exp{\left[-i(\bar{k}_{\bar{a}}+k_{\bar{a}})x\right ] } \nonumber \\ 
		&+&
		\frac{e}{2h}\sum_{b=1}^{\hat{N}}
		\left(\sqrt{\frac{\bar{k}_b}{k_b}}+
		\sqrt{\frac{k_b}{\bar{k}_b}}\right)t^{*}_{ba}
		{\bar r}'_{b\bar{a}}
		\exp{\left[i(\bar{k}_b-k_b)x\right]}
		.
	\end{eqnarray}
\end{widetext}
While the matrix $t$ is taken at the energy $\varepsilon$, the matrices $\bar{t}$ and $\bar{r}'$ are evaluated at the energy $\bar{\varepsilon}$. The summations in \eqref{detailed2} is up to $\hat{N}=min(N(\varepsilon),N(\bepsilon))$.

According to the results in Sec. \ref{energyintegrals}, the first term of $\l[j(\br)\r]_{\ba a}^{12}(\varepsilon,\bar{\varepsilon})$ plugged in \eqref{eq:j1} leads to a vanishing energy integral.
Therefore, only $\l[j(\br)\r]_{\ba a}^{11}(\varepsilon,\bar{\varepsilon})$ and the second term of $\l[j(\br)\r]_{\ba a}^{12}(\varepsilon,\bar{\varepsilon})$ contribute to  $\delta ^{(1)}J_{1\varepsilon a}(\br)$. 

Using Eq. \eqref{i1pappa}, we find that the current resulting from a given mode $a$ in lead 1 at energy $\varepsilon$ in first-order in the tip potential is
\begin{equation}
	\label{eq:I1a}
	I_{1,\varepsilon,a}^{(1)} = \frac{e}{\hbar}\, \mathrm{Im}\left\lbrace 
	t^{\dagger}t\ \mathcal{V}^{11}
	+ t^{\dagger}r'\ \mathcal{V}^{21}
	\right\rbrace_{aa}  \, ,
\end{equation}
where $\mathcal{V}^{l\bl}_{a\bar{a}}=\left[V_{\rm{T}}\right]^{l\bl}_{a\bar{a}}(\varepsilon,\varepsilon)$.
Summing over all the modes we obtain the total current as
\begin{equation}
	\label{eq:I1}
	I_{1,\varepsilon}^{(1)} = \frac{e}{\hbar}\, \mathrm{Im}(\mathrm{Tr}\left\lbrace 
	t^{\dagger}t\ \mathcal{V}^{11}
	+ t^{\dagger}r'\ \mathcal{V}^{21}
	\right\rbrace)  \, .
\end{equation}
In the linear response regime the zero temperature conductance is obtained by differentiating the total current with respect to the applied voltage, that is

\begin{equation}
	\label{eq:G1}
	G^{(1)} = \frac{2e}{\Delta \mu}(\Delta \mu I_{1,\EF}^{(1)} ),
\end{equation}
with $\Delta \mu $ the chemical potential difference between the two probes. The  $\Delta \mu$ factor in the numerator results from the fact that at zero temperature only the Fermi energy current contributes to the transport and therefore the integration between the two chemical potentials reduces to a multiplication by $\Delta \mu$.
This leads to the dimensionless conductance (in units of $\frac{2e^2}{h}$)   \cite{gorini2013,jalabert2010} 
\begin{equation}
	\label{eq:g11}
	g^{(1)} = 4 \pi \mathrm{Im}\left\lbrace \mathrm{Tr}\l[ 
	t^{\dagger}t\ \mathcal{V}^{11}
	+t^{\dagger}r'\ \mathcal{V}^{21}\r]
	\right\rbrace  \, .
\end{equation}

Yet, plugging the expression of ${\Psi}_{l\varepsilon a} ^{(1)}(\br)$ in (\ref{J2beta1}), the second order contribution to the current density  can be brought in the following form
\begin{widetext}
	\begin{equation}\label{betalikej}
		\delta ^{(2)}J_{1\varepsilon a} =
		\frac{e\hbar}{\Me}  
		\sum_{\bl,\bbl=1}^{2} \mathrm{Im}\left\lbrace
		\int_{\tepsilon_1}^{\infty} 
		\frac{\dif\bepsilon}{\varepsilon^{-}-\bepsilon} 
		\int_{\tepsilon_1}^{\infty} 
		\frac{\dif\bbepsilon}{\varepsilon^{+}-\bbepsilon} 
		\sum_{\ba,\bba}
		\left[\VT\right]^{l \bl}_{a \ba}(\varepsilon,\bepsilon)
		\left[j_{1/2}\right]_{\ba \bba}^{\bl, \bbl}(\bepsilon,\bbepsilon)\
		\left[\VT\right]^{\bbl l}_{\bba a}(\bbepsilon,\varepsilon)
		\right\rbrace ,
	\end{equation}
\end{widetext}
where 
\begin{equation}\label{j1/2}
	\l[j_{1/2}\r]_{\ba a}^{\bl l}(\bar{\varepsilon},\varepsilon)=
	\frac{e\hbar}{2i\Me}
	\int
	\l[
	\Psi_{\bar{l},\bar{\varepsilon},\bar{a}}^{(0)*}(\br) 
	\ {\partial _x}\Psi_{l,\varepsilon,a}^{(0)}(\br)  
	\r] \dif y.
\end{equation}

The matrix elements of $[j_{1/2}]$  are straightforwardly obtained by plugging the scattering wave-functions \eqref{allscats} in  Eq. \eqref{j1/2}, they read

\begin{widetext}
\begin{eqnarray}
	\label{eq:jonehalf}
	{[j_{1/2}]}^{11}_{a \ba}(\varepsilon,\bepsilon) &=& \frac{e}{2h}  
	\sum_{b=1}^{\hat{N}} \sqrt{\frac{\bk_b}{k_b}} \
	t^{*}_{ba} {\bar t}^{\phantom{*}}_{b \ba} \
	e^{i(\bk_{b}-k_{b})x } \, , \nonumber \\
	{[j_{1/2}]}^{12}_{a \ba}(\varepsilon,\bepsilon) &=& \frac{e}{2h}
	\left\lbrace -
	\sqrt{\frac{\bk_{\ba}}{k_{\ba}}} \ t^{*}_{\ba a} \
	e^{-i(\bk_{\ba}+k_{\ba})x }
	+ \sum_{b=1}^{\hat{N}} \sqrt{\frac{\bk_b}{k_b}} \ t^{*}_{ba}{\bar r}'_{b\ba} \
	e^{i(\bk_{b}-k^-_b)x}
	\right\rbrace , \nonumber \\
	{[j_{1/2}]}^{21}_{a \ba}(\varepsilon,\bepsilon) &=& \frac{e}{2h}
	\left\lbrace 
	\sqrt{\frac{\bk_{a}}{k_{a}}} \ {\bar t}^{*}_{a \ba} \
	e^{i(\bk_{a}+k_{a})x }
	+ \sum_{b=1}^{\hat{N}} \sqrt{\frac{\bk_b}{k_b}} \ r'^{*}_{ba} {\bar t}_{b \ba} \
	e^{i(\bk_{b}-k_b)x}
	\right\rbrace ,\nonumber \\
	{[j_{1/2}]}^{22}_{a \ba}(\varepsilon,\bepsilon) &=& \frac{e}{2h}
	\left\lbrace
	-\delta_{a \ba} \ \sqrt{\frac{\bk_{a}}{k_{a}}} \
	e^{-i(\bk_{\ba}-k_{a})x } +
	\sqrt{\frac{\bk_{a}}{k_{a}}} \ {\bar r}'_{a \ba}
	e^{i(\bk_{a}+k_{a})x} 
	\right. \nonumber
	\\
	& & 
	-\left.
	\sqrt{\frac{\bk_{\ba}}{k_{a}}} \ r'^{*}_{\ba a} \
	e^{-i(\bk_{\ba}+k_{\ba})x } + 
	\sum_{b=1}^{\hat{N}} \sqrt{\frac{\bk_b}{k_b}} \ r'^{*}_{ba} {\bar r}'_{b \ba} \
	e^{i(\bk_{b}-k_b)x}
	\right\rbrace .\nonumber\\
	\ 
\end{eqnarray}
\end{widetext}

The first terms of ${[j_{1/2}]}^{12}$ and ${[j_{1/2}]}^{21}$ leads respectively to energy integrals of types $I_2^{(-,+)}$ and $I_2^{(+,+)}$, and therefore vanish according to \eqref{i2pmappa}, stating that only integrals of type $I_2^{(+,-)}$ are non zero.
Similarly, the first three terms of ${[j_{1/2}]}^{22}$ leads to vanishing energy integrals, respectively of types $I_2^{(-,+)}$, $I_2^{(+,+)}$ and $I_2^{(-,+)}$.
By consequence, only the last  terms of ${[j_{1/2}]}^{12}$ , ${[j_{1/2}]}^{21}$ and ${[j_{1/2}]}^{22}$ in addition to ${[j_{1/2}]}^{11}$ have to be considered.
The energy integrals due to these terms  are obtained using \eqref{i2pmappa}.
They red respectively,
\begin{equation}
	4\pi^2\frac{e}{2h}(\mathcal{V}^{11}t^{\dagger}t\mathcal{V}^{11})_{aa},
\end{equation}

\begin{equation}
	4\pi^2\frac{e}{2h}(\mathcal{V}^{11}t^{\dagger}r'\mathcal{V}^{21})_{aa},
\end{equation}

\begin{equation}
	4\pi^2\frac{e}{2h}(\mathcal{V}^{12}t^{\dagger}t\mathcal{V}^{11})_{aa},
\end{equation}
and 
\begin{equation}
	4\pi^2\frac{e}{2h}(\mathcal{V}^{12}t^{\dagger}t\mathcal{V}^{21})_{aa}.
\end{equation}
The resulting dimensionless conductance correction is obtained as a sum of these terms \cite{jalabert2010,gorini2013,szewcthesis,lythesis}. It reads
\begin{equation}
	\label{eq:g2beta}
	{g^{(2)} = 4\pi^2  \mathrm{Tr}\left\lbrace 
		{\rm {Re}}[\mathcal{V}^{11}t^\dagger t \mathcal{V}^{11}+2\mathcal{V}^{11} t^\dagger r' \mathcal{V}^{21}+\mathcal{V}^{12}r'^\dagger r'\mathcal{V}^{21}]
		\right\rbrace  \,}.
\end{equation}

Eq. \eqref{eq:g2beta}, is indeed valid for any tip potential. However, if a delta tip is assumed, the matrix elements can be further simplified leading to the following expression.
\begin{equation}
	\label{eq:g2beta-factored}
	{g^{(2)} = 4\pi^2  \mathrm{Tr}\left\lbrace 
		{\rm {Re}}[\mathcal{V}^{11}(t^\dagger t \mathcal{V}^{11}+2 t^\dagger r' \mathcal{V}^{21}+r'^\dagger r'\mathcal{V}^{22})]
		\right\rbrace  \,}.
\end{equation}




\bibliography{sgmrefs}
\end{document}